\documentclass[aps,prb,reprint,groupedadress,showpacs]{revtex4-1}
\usepackage{amssymb}
\usepackage{amsmath}
\usepackage{graphicx}
\usepackage{color}
\begin{document}
\title{Interpretation of the Dispersion of the Electron States of High-$T_{c}$ Cuprate Superconductors Based on the Theory of Topological Resonance}
\author{W.Winkler}
\affiliation{Laboratory for Materials, Wackenbergstr.\,84-88, 13156\,Berlin, Germany}
\date{\today}
\begin{abstract}
The unusual dispersion of electron states in antinodal direction for $n_{h}\simeq\frac{1}{8}$ holes/copper is explained. In photoelectron excitations, transformations of the $\mathbf{k}$ space as a consequence of symmetry breaking proves to be an exceptional phenomenon distinctly reflected by the experiments. Paired photoelectron emissions are predicted that enable a new energetic upward shift termed as \emph{pair shake-up effect} which explains the experimentally observed sharp transition between non-dispersive and dispersive behaviour. The disappearance of antinodal states for $\mathbf{k}_{||}\epsilon[0,\pm\frac{\pi}{2}]$ is explained.
\end{abstract}
\maketitle
\section{Introduction\label{secnj1}}
$\indent$The electron dispersion in high-$T_{c}$
cuprates is one unresolved problem and is not consistently explained by theory.\cite{rjarp20} Considering
single layer systems, the electronic states near the Fermi level are completely dispersionless in antinodal
direction for $0.75\pi$$\leq$$\mathbf{k}_{||}$$<$$1.25\pi$  but strongly disperse for
$0.5\pi$$\leq$$\mathbf{k}_{||}$$<0.75\pi$. This  behaviour was experimentally observed by angle-resolved
photoemission spectroscopy (ARPES) of  $(\mathrm{La_{1.28}Nd_{0.6}Sr_{0.12})CuO_{4}}$.\cite{rjarp5} For the
parent compound $(\mathrm{La_{2-x}Sr_{x})CuO_{4}}$ the same dispersion behaviour was found for $0.1$$\leq$ x
$\leq$0.3 and even for strongly reduced hole densities up to $x=0.05$ dispersionless states were observed, but
which are shifted below the Fermi level.\cite{rjarp6,rjarp7} A similar antinodal dispersion can be observed in
$(\mathrm{Bi_{1.80}Pb_{0.38}Sr_{2.01})CuO_{6-\delta}}$.\cite{rjarp8} It is noticeable that near the Fermi level
the antinodal states obviously disappear completely for 0$\leq$$\mathbf{k}_{||}$$<$$0.5\pi$.\cite{rjarp5}$^{-}$\cite{rjarp8} The dispersionless
states in antinodal direction $k_{||}$ exist for perpendicular wave vector components
of 0$\leq {k}_{\bot}$$<$$\pm0.25\pi$. Within this range the states are dispersionless
also in ${k}_{\bot}$ direction for nominal hole densities
$n_{h}(\equiv x)=0.12-0.3$ holes/copper.\cite{rjarp5}$^{-}$\cite{rjarp7} These extraordinary
experimental findings will be proven here by a new theoretical concept, termed as
topological resonance (TR) theory\cite{rjarp1,rjarp2}, which defines new quantum states in high-$T_{c}$ cuprates.
\section{The electronic basis states of the $\mathbf{CuO_{2}}$ plane\label{secnj2}}
\begin{figure}[tpb]
\centerline{\includegraphics[width=0.48\textwidth]{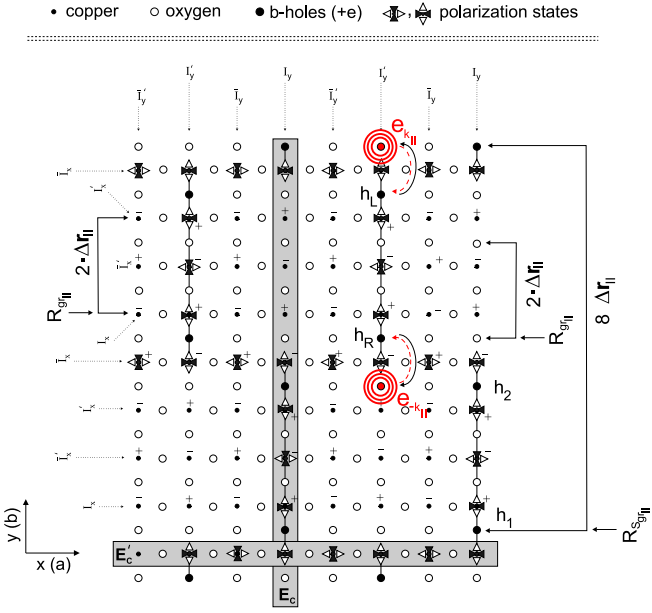}}
\caption{\label{fjarp1}
Topological hole configuration of a $\mathrm{CuO_{2}}$ plane for a nominal hole concentration
$n_{h}$=$\frac{1}{8}$ holes/copper forming an ordered topological hole state over oxygen sites, termed as b-hole state.\cite{rjarp1} The valence bands, usually being two-dimensional, are renormalized into two perpendicular arrays of one-dimensional bands, along the \emph{x} coordinate
$\{I_{x},I'_{x},\bar{I}_{x},\bar{I}'_{x}\}$ and the \emph{y} coordinate $\{I_{y},I'_{y},\bar{I}_{y},\bar{I}'_{y}\}$.\cite{rjarp1} The +,$-$ signs indicate two different copper sites within every one-dimensional band resulting from renormalized electronic basis states, CBF states, which define new lattice periodicities of $2a$ and $2b$ in $x$ and $y$ direction, respectively. The b-hole topology results from attractive b-hole$-$b-hole pair couplings $E_{C}<0$\,\cite{rjarp1} depicted, e.g., within the shadowed one-dimensional bands. The b-hole state has a periodicity of $8b$ in $y$ direction, $4a$ in $x$ direction, and within every one-dimensional band two periodic b-hole substructures ($h_{1}$,$h_{2}$) exist, shifted to each other by $3b$. These two b-hole substructures create two separate electron channels in which individual $\mathbf{k}_{||}$, $-\mathbf{k}_{||}$ states exist being intrinsically paired.\cite{rjarp2} In photoelectron emission, two paired electrons $e_{\mathbf{k}_{||}}$, $e_{-\mathbf{k}_{||}}$ are emitted, for example, from the red circled lattice sites. An additional electron$-$b-hole pair exchange may occur corresponding to the red (dotted) and black arrows creating an additional pair shake-up effect.}
\end{figure}

$\indent$Instabilities of charge and spin density orders in the $\mathrm{CuO_{2}}$ planes are often discussed
in literature.\cite{rjarp9'}$^{-}$\cite{rjarp12} Antiferromagnetic correlations and coulomb frustrated phase separations
 are assumed to be the generic features of these highly correlated electronic systems. An antiferromagnetic Mott
 insulator is postulated as initial state for the half filled bands. According to this assumption, doping of holes
 will lead to a Mott insulator-conductor transition which leaves behind subtle couplings between electronic and spin
 dynamics on a microscopic scale. In a more stringent ab initio approach, however, we could show that the initial
 electronic states are already symmetry broken in various ways leading to new ordered  macroscopic electronic
 structures but which are still dynamic (TR state).\cite{rjarp1,rjarp2} Basically, the two-dimensionality of
 the energetically highest valence bands is removed in favour to two arrays of one-dimensional
 bands \{$I_{x},I'_{x},\bar{I}_{x},\bar{I}'_{x}\}$ and \{$I_{y},I'_{y},\bar{I}_{y},\bar{I}'_{y}$\} polarized
 along the \emph{x} and \emph{y} direction, respectively (Fig.\,\ref{fjarp1}). Highly exceptional is that the
 electronic renormalizations lead to two different copper sites, $-$ and $+$ (Fig.\,\ref{fjarp1}), which
 define two disjoint basis representations of the one-dimensional periodic local states $u_{\mathbf{k}_{||}}$ having
 a periodicity of $2\Delta\mathbf{r}_{||}$=$2a$ or $2\Delta\mathbf{r}_{||}$=$2b$. These disjoint basis representations
 span two separated Hilbert spaces $\stackrel{[-]}{\mathcal{H}}_{I_{i}, -\frac{\Delta\mathbf{r_{||}}}{2}}$ and $\stackrel{[+]}{\mathcal{H}}_{I_{i},  +\frac{\Delta\mathbf{r_{||}}}{2}}$ related to the $-$ and $+$ lattice sites, respectively. In local space, these separated Hilbert spaces  define two disjoint coordinate systems with a definite relative coordinate $\Delta\mathbf{r}_{||}$=$a$ or $\Delta\mathbf{r}_{||}$=$b$ to each other. For the static electronic ground state in Fig.\,\ref{fjarp1}, only the Hilbert spaces $\stackrel{[-]}{\mathcal{H}}_{I_{i}, -\frac{\Delta\mathbf{r_{||}}}{2}}$ are populated in the hole undoped state, so that a charge and bonding fluctuation state (CBF) with the period 2\emph{a}, 2\emph{b} is established, collinearly with the $-$ and $+$ signs in Fig.\,\ref{fjarp1}.\\
$\indent$Doping of holes results in additional symmetry breaking leading to the formation of a strongly
ordered topological hole structure for a nominal hole density of $n_{h}$=1/8, termed as b-hole state. Every particular
topological hole state is assigned to a given oxygen site. As a result, a periodic hole order with the periodicity
of $8b$ along the one-dimensional bands $I_{y},I'_{y}$ in $y$ direction and of $4a$ in $x$ direction is
established (Fig.\,\ref{fjarp1}). Asymmetric hole doping on particular oxygen sites requires horizontal electronic
transformations in real space with the result that every local state $u_{i}$ related to a given copper site $i$ is
split up into a left and right part which are separately associated to the neighbouring left and right oxygen atom
in a O-Cu-O bonding unit, respectively, subsequently abbreviated by $L$ and $R$.\cite{rjarp1,rjarp3} In consequence, the initial Hilbert spaces are additionally
split up according to $\stackrel{[-]}{\mathcal{H}}_{I_{i}, -\frac{\Delta\mathbf{r_{||}}}{2}}\rightarrow\stackrel{[-]}{\widetilde{\mathcal{H}}}_{I_{i}, -\Delta\mathbf{r_{||}}},  \stackrel{[-]}{\widetilde{\mathcal{H}}}_{I_{i}, 0}$ and $\stackrel{[+]}{\mathcal{H}}_{I_{i}, +\frac{\Delta\mathbf{r_{||}}}{2}}\rightarrow \stackrel{[+]}{\widetilde{\mathcal{H}}}_{I_{i}, 0}, \stackrel{[+]}{\widetilde{\mathcal{H}}}_{I_{i}, +\Delta\mathbf{r_{||}} \equiv -\Delta\mathbf{r_{||}}}$,
but interference parts between the split local states are still preserved.\cite{rjarp3} The accompanied coordinate
transformations $-\frac{\Delta\mathbf{r_{||}}}{2}\rightarrow-\Delta\mathbf{r_{||}},0$
and $+\frac{\Delta\mathbf{r_{||}}}{2} \rightarrow 0,+\Delta\mathbf{r_{||}}$ reflect the transformations
of the coordinate systems from the copper to the oxygen sites. Thus, every split Hilbert space creates two
separated electron channels which belong to different oxygen sites, which is reflected, for example, by the hole
positions $h_{1}$ and $h_{2}$ in Fig.\,\ref{fjarp1}. As previously shown,\cite{rjarp2} the one-dimensional Bloch
states with impulse vectors $-\mathbf{k}_{||}$ and $\mathbf{k}_{||}$ separately belong to the two electron channels
and form intrinsic electron pair states. The quantum states within every electron channel exhibit a periodicity
of $8b$, and the two channels are shifted to each other by $3b$. The static CBF state in Fig.\,\ref{fjarp1}, however,
is not the ground state, but the state which results from a dynamic superposition of the CBF state with its mirror
symmetric state $\mathrm{\overline{CBF}}$, termed as DCBF state (TR).\cite{rjarp2}  The b-hole state
in Fig.\,\ref{fjarp1} is also dynamic in $y$ as well as in $x$ direction which leads to various
additional topological resonance states.\cite{rjarp2}
\section{Photoexcitations under disjoint basis states\label{secnj3}}
$\indent$The discussion of the electron dispersion is related to photoelectron emission (ARPES) data, the three
step model being taken as a basis.\cite{rjarp20} However, the first step that implies optical excitations in the bulk is distinctly different. First of all, the collective electronic b-hole state bears generally a collective electronic response. In addition, the formation of disjoint coordinate systems and
their internal transformations, as discussed above, constitutes extraordinary aspects concerning optical excitations
of the electrons in the bulk. Hence, the subsequent discussions are initially focused on electronic transitions from the
ground to the excited states. In order to comprehend the problem, the Bloch functions have to be extended by
\emph{coordinate} \emph{state} \emph{functions} for the ground and the excited state according to
\begin{eqnarray}\label{ejap1}
\psi'_{\mathbf{k}_{gr}}&=&e^{i\mathbf{k}\cdot\mathbf{r}}\cdot
e^{i\mathbf{G}_{{gr}}^{cr}\cdot\mathbf{r}_{{gr}}^{cr}}u_{\mathbf{k}_{gr}}(\mathbf{r})
= \psi_{\mathbf{k}}\cdot\psi_{\mathbf{G}_{{gr}}^{cr}}\,\in\, \mathcal{H}_{\mathbf{G}^{cr}_{{gr}}}\hspace{0.5cm}\\
\label{ejap2}
\psi'_{\mathbf{k}_{ex}}&=&e^{i\mathbf{k}\cdot\mathbf{r}}\cdot
e^{i\mathbf{G}_{{ex}}^{cr}\cdot\mathbf{r}_{{ex}}^{cr}}u_{\mathbf{k}_{ex}}(\mathbf{r})
= \psi_{\mathbf{k}}\cdot\psi_{\mathbf{G}_{{ex}}^{cr}}\,\in\,\mathcal{H}_{\mathbf{G}^{cr}_{{ex}}}
\end{eqnarray}
\begin{eqnarray}\label{ejap3}
\hspace{-1.6cm}\mathrm{with} \hspace{0.6cm}e^{i\mathbf{G}_{{gr}}^{cr}\cdot\mathbf{r}_{{gr}}^{cr}}\equiv 1 \hspace{0.4cm}\mathrm{and}  \hspace{0.4cm} e^{i\mathbf{G}_{{ex}}^{cr}\cdot\mathbf{r}_{{ex}}^{cr}}\equiv 1,
\end{eqnarray}
where $\psi_{\mathbf{G}_{{gr}}^{cr}}$ and $\psi_{\mathbf{G}_{{ex}}^{cr}}$ are coordinate state functions and  $\mathbf{G}_{{gr}}^{cr}$, $\mathbf{G}_{{ex}}^{cr}$ the corresponding state vectors that define the equivalence of lattice sites $\mathbf{r}_{{gr}}^{cr}$ and $\mathbf{r}_{{ex}}^{cr}$ for assigning the coordinate system in the ground and the excited state, respectively. $\mathcal{H}_{\mathbf{G}^{cr}_{{gr}}}$,  $\mathcal{H}_{\mathbf{G}^{cr}_{{ex}}}$ are the corresponding Hilbert spaces. Usually, one can assume $\psi_{\mathbf{G}_{{gr}}^{cr}} \equiv \psi_{\mathbf{G}_{{ex}}^{cr}}$ making coordinate state functions meaningless. However, the one-dimensional valence quantum states of the $\mathrm{CuO_{2}}$ planes are symmetry broken resulting in the general wave function for the one-dimensional b-hole states within the particular bands $I_{y}$, $I'_{y}$ of Figs.\,\ref{fjarp1} according to:
\begin{eqnarray}\label{ejap4}
&&\hspace{-0.8cm}\stackrel{[-]}{\psi'}_{\mathbf{k}_{||_{gr}}}=
e^{i\mathbf{k}_{||}\cdot\stackrel{[-]}{\mathbf{r}}\hspace{-0.0cm}'\hspace{-0.25cm}_{1}} \cdot e^{i\stackrel{[-]}{\mathbf{G}}{\hspace{-0.0cm}^{cr}_{{gr}_{||}}}\cdot\mathbf{R}_{gr_{||}}(t')} \cdot
e^{i\frac{1}{4}\stackrel{[-]}{\mathbf{G}}{\hspace{-0.0cm}^{cr}_{{gr}_{||}}}\cdot\mathbf{R}_{S_{gr_{||}}}(t')} \\
&&\hspace{0.5cm}\cdot \,e^{i\stackrel{[-]}{\mathbf{G}}{\hspace{-0.0cm}^{cr}_{{gr}_{||}}}\cdot \mathbf{\stackrel{[-]}{r}}\hspace{-0.0cm}^{'cr}_{gr_{||}}}\cdot
e^{i\frac{1}{4}\stackrel{[-]}{\mathbf{G}}{\hspace{-0.0cm}^{cr}_{{gr}_{||}}}\cdot \stackrel{[-]}{\mathbf{r}}\hspace{-0.0cm}^{'cr}_{S_{gr_{||}}}} \cdot \stackrel{[-]}{u}_{R,\mathbf{k}_{||_gr}}(\stackrel{[-]}{\mathbf{r}}\hspace{-0.05cm}'\hspace{-0.25cm}_{1}\,),\nonumber
\end{eqnarray}
for a $-$ lattice site, with ${\stackrel{[-]}{\mathbf{r}}\hspace{-0.1cm}'\hspace{-0.2cm}_{1}}={\stackrel{[-]}{\mathbf{r}}\hspace{-0.2cm}_{1}}-\mathbf{R}_{gr_{||}}(t')-\mathbf{R}_{S_{gr_{||}}}(t'))$, $\mathbf{\stackrel{[-]}{r}}\hspace{-0.0cm}^{'cr}_{gr_{||}}=\mathbf{\stackrel{[-]}{r}}\hspace{-0.0cm}^{cr}_{gr_{||}}-\mathbf{R}_{gr_{||}}(t')$ and $\stackrel{[-]}{\mathbf{r}}\hspace{-0.2cm}^{'cr}_{S_{gr_{||}}}={\stackrel{[-]}{\mathbf{r}}\hspace{-0.1cm}^{cr}_{S_{gr_{||}}}-\mathbf{R}_{S_{gr_{||}}}(t')}$. Eq.\,(\ref{ejap4}) includes two symmetry broken coordinate state functions. The state vector
${\stackrel{[-]}{\mathbf{G}}{\hspace{-0.1cm}}{^{cr}_{{gr}_{||}}}}$=$\frac{1}{2}\mathbf{G}^{cr}_{{o}_{||}}$$\equiv$$\frac{1}{2}\frac{2\pi}{a}$\,$\mathrm{OR}$\,$\equiv$$\frac{1}{2}\frac{2\pi}{b}$
in direction of the one-dimensional bands describes the coordinate relative state arising from the symmetry
broken CBF state with the particular phase
$\stackrel{[-]}{\mathbf{G}}{\hspace{-0.13cm}{^{cr}_{{gr}_{||}}}}$$\cdot\mathbf{R}_{gr_{||}}(t')$ relative to the
coordinate system of the lattice. The state vector
${\frac{1}{4}{\hspace{-0.1cm}}\stackrel{[-]}{\mathbf{G}}{\hspace{-0.1cm}}{^{cr}_{{gr}_{||}}}}$=$\frac{1}{8}\mathbf{G}^{cr}_{{o}_{||}}$
is related to the additionally symmetry broken b-hole superstructure with the particular phase
$\frac{1}{4}{\hspace{-0.1cm}}\stackrel{[-]}{\mathbf{G}}{\hspace{-0.1cm}^{cr}_{{gr}_{||}}}$$\cdot\mathbf{R}_{S_{gr_{||}}}(t')$
within the given CBF state, defining a coordinate relative state of the b-hole state relative to the coordinate
system of the CBF state. The time $t'$ represents an internal quantum clock-time $t'=n\cdot\tau_{CBF}$, with
$\tau_{CBF}\simeq1.5\cdot10^{-14}s$ being the lifetime of the given CBF state.\cite{rjarp1,rjarp2} The symmetry
breakings of the renormalized valence ground states are caused by a collective electronic behaviour of the
populated states whereas the unpopulated non-renormalized excited states are initially non-symmetry broken.
Thus, a non-symmetry broken coordinate state function with a coordinate state vector
$\mathbf{G}^{cr}_{{ex}_{||}}$=$\mathbf{G}^{cr}_{{o}_{||}}$ has to be assumed. For the step of optical
excitations in the bulk, electronic excitations from the ground state to the excited states have to be
considered. The corresponding transition matrix elements require, however, a common coordinate system here,
defined by the coordinate state vector $\mathbf{G}^{cr}_{{ex}_{||}}$ of the excited states. Therefore, the
initial coordinate state function
$e^{i\stackrel{[-]}{\mathbf{G}}{\hspace{-0.0cm}^{cr}_{{gr}_{||}}}\cdot \mathbf{\stackrel{[-]}{r}}\hspace{-0.0cm}^{'cr}_{gr_{||}}}\cdot
e^{i\frac{1}{4}\stackrel{[-]}{\mathbf{G}}{\hspace{-0.0cm}^{cr}_{{gr}_{||}}}\cdot \stackrel{[-]}{\mathbf{r}}\hspace{-0.0cm}^{'cr}_{S_{gr_{||}}}}$
in Eq.\,(\ref{ejap4}) will be replaced by the non-symmetry broken
coordinate state function
$\psi_{\mathbf{G}_{{ex_{||}}}^{cr}}$=$e^{i\stackrel{}{\mathbf{G}}{\hspace{-0.0cm}^{cr}_{{ex}_{||}}}\cdot\mathbf{{r}}^{'cr}_{ex_{||}}}$
henceforth with the probabilistic variable
$\mathbf{{r}}^{'cr}_{ex_{||}}=\mathbf{\stackrel{[-]}{r}}\hspace{-0.0cm}^{'cr}_{gr_{||}}+\stackrel{[-]}{\mathbf{r}}\hspace{-0.1cm}^{'cr}_{S_{gr_{||}}}-\mathbf{{r}}^{cr}_{ex_{||}}$
which is defined for all lattice sites within a given one-dimensional band. The two phases
${\stackrel{[-]}{\mathbf{G}}{\hspace{-0.1cm}^{cr}_{{gr}_{||}}}}$$\cdot\mathbf{R}_{gr_{||}}(t')$ and
${\frac{1}{4}\stackrel{[-]}{\mathbf{G}}{\hspace{-0.1cm}^{cr}_{{gr}_{||}}}}$$\cdot\mathbf{R}_{S_{gr_{||}}}(t')$
are conserved but the new excited state variable $\mathbf{r}^{cr}_{ex_{||}}$ transforms the coordinates
according to
$\mathbf{R}_{gr_{||}}(t')\rightarrow\mathbf{{r}}'_{{||}}=\mathbf{R}_{gr_{||}}(t')-\mathbf{r}^{cr}_{ex_{||}}$
and
$\mathbf{R}_{gr_{||}}(\mathbf{r}_{S_{||}})(t')\rightarrow\mathbf{{r}}'_{S_{{||}}}=\mathbf{R}_{gr_{||}}(\mathbf{r}_{S_{||}})(t')-\mathbf{r}^{cr}_{ex_{||}}$
with $\mathbf{{r}}'_{{||}}$ and $\mathbf{{r}}'_{S_{{||}}}$ representing new probabilistic relative
coordinates with the actual positions $\mathbf{R}_{gr_{||}}(t')$ and $\mathbf{R}_{S_{gr_{||}}}(t')$ being
maintained. Thus, the initially stationary phases
${\stackrel{[-]}{\mathbf{G}}{\hspace{-0.1cm}^{cr}_{{gr}_{||}}}}$$\cdot\mathbf{R}_{gr_{||}}(t')$ and
${\frac{1}{4}\stackrel{[-]}{\mathbf{G}}{\hspace{-0.1cm}^{cr}_{{gr}_{||}}}}$$\cdot\mathbf{R}_{S_{gr_{||}}}(t')$, i.e. stationary at the given quantum clock-time $t'$,
are converted into a probabilistic quantum state variable, respectively, at every time. This leads to state
functions according to:
\begin{eqnarray}
 \label{ejap4'}\hspace{-0.3cm}\psi'_{\mathbf{k}_{||_{ex}}}=e^{i\mathbf{k}_{||}\cdot\mathbf{{r}}_{1}'}\hspace{-0.1cm}\cdot e^{i{\mathbf{k}_{{CBF}_{||}}}\cdot\mathbf{{r}}'_{{||}}}\cdot
 e^{i{\mathbf{k}_{{S}_{||}}}\cdot \mathbf{{r}}'_{S_{{||}}}}\cdot u_{\mathbf{k}_{||_ex}}(\mathbf{{r}}_{1}')\,\psi_{\mathbf{G}_{{ex_{||}}}^{cr}}\nonumber\\
 &&
\end{eqnarray}
with $\mathbf{k}_{CBF_{||}}$$\equiv$$\frac{1}{2}\mathbf{G}^{cr}_{{o}_{||}}$ being the wave vector of the CBF state and $\mathbf{k}_{S_{||}}$$\equiv$$\frac{1}{8}\mathbf{G}^{cr}_{{o}_{||}}$ the wave vector of the b-hole state.\\
$\indent$The transformation of Eq.\,(\ref{ejap4}) to Eq.\,(\ref{ejap4'}), however, will only be allowed, if a
free transformation of the renormalized local states
$\stackrel{[-]}{u}_{R,\mathbf{k}_{||_gr}}\hspace{-0.2cm}(\stackrel{[-]}{\mathbf{r}}\hspace{-0.15cm}'\hspace{-0.2cm}_{1})$ to the non-renormalized local states
$u_{\mathbf{k}_{||_ex}}(\mathbf{{r}}\hspace{-0.0cm}'\hspace{-0.1cm}_{1})$ is possible. Usually, in hole free states like
$\bar{I}_{y},\bar{I}'_{y}$, such unrestricted transitions are not allowed because the coordinate systems of
$\stackrel{[-]}{\mathcal{H}}_{I_{i}, -\frac{\Delta\mathbf{r_{||}}}{2}}$ and
$\stackrel{[+]}{\mathcal{H}}_{I_{i},+\frac{\Delta\mathbf{r_{||}}}{2}}$ are disjoint. For b-hole states within
$I_{y},I'_{y}$, the transformed coordinate systems of the Hilbert spaces permit, however, quantum states which
are defined within the linear span of $\stackrel{[-]}{\widetilde{\mathcal{H}}}_{I_{i},
0}$\,$\cup$\,$\stackrel{[+]}{\widetilde{\mathcal{H}}}_{I_{i}, 0}$ and
$\stackrel{[-]}{\widetilde{\mathcal{H}}}_{I_{i}-\Delta\mathbf{r_{||}}}$\,$\cup$\,$\stackrel{[+]}{\widetilde{\mathcal{H}}}_{I_{i}, +\Delta\mathbf{r_{||}}\equiv-\Delta\mathbf{r_{||}}}$, with conserved time functions of $\mathbf{R}_{gr_{||}}(t')$ and $\mathbf{R}_{S_{gr_{||}}}(t')$.\cite{rjarp3} This allows unrestricted probabilistic transitions between Eqs.\,(\ref{ejap4}) and (\ref{ejap4'}) under conserving the time-quantization of the states which is defined by $t'$.
\section{Paired photoelectron emissions\label{secnj4}}
\begin{figure}[tpb]
\centerline{\includegraphics[width=0.45\textwidth]{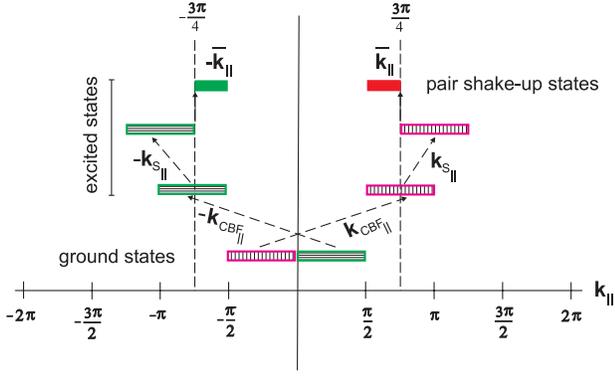}}
\caption{\label{fjarp2}
Transformation of a symmetry broken valence band (ground state) into a non-symmetry broken excited band state
for $-\mathbf{k}_{{||}}$, $\mathbf{k}_{{||}}$ paired electron channels. The two electron channels are
pairwise shifted by the CBF state vectors $\pm\mathbf{k}_{{CBF}_{||}}$ and the b-hole state
vectors $\pm\mathbf{k}_{S_{||}}$. Additional electron$-$b-hole pair exchange leads to the
depicted pair shake-up states.}
\end{figure}

\begin{figure}[tpb]
\vspace{-0.0cm}\centerline{\includegraphics[width=0.47\textwidth]{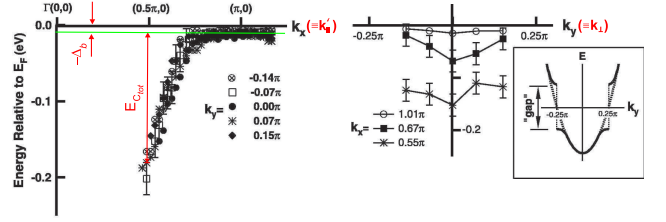}}
\caption{\label{fjarp3}
Experimental dispersion of $(\mathrm{La_{1.28}Nd_{0.6}Sr_{0.12})CuO_{4}}$  taken from Ref.\,[2] (leading-edge). Reprinted with permission from AAAS. We identify the small shift below the Fermi level by $-\Delta_{b}$, and the linear dispersion results from the pair shake-up effect with a maximum value of $E_{C_{tot}}$ at $\bar{k}_{||}=\frac{\pi}{2}$. States for $k_{x}(\textcolor{red}{\equiv k'_{||}})\epsilon[0,\frac{\pi}{2}]$ are disappeared. Near the Fermi level, the states $k_{y}(\textcolor{red}{\equiv k_{\bot}})$ are dispersionless.}
\end{figure}

$\indent$According to Eq.(\ref{ejap4'}) three simultaneous vertical transitions occur for b-hole states in photoelectron excitations: the ordinary $\mathbf{k}_{||}$ state excitation, the transition related to the CBF relative state with $\mathbf{k}_{CBF_{||}}$ and the transition corresponding to the b-hole relative state with $\mathbf{k}_{S_{||}}$. In addition, the excitations occur pairwise with $\mathbf{k}_{{||}}$, $-\mathbf{k}_{{||}}$ impulse vectors which are separately assigned to the two electron channels in Fig.\,\ref{fjarp1} because only paired delocalized electron states exist within the bands $I_{y}$, $I'_{y}$ in Fig.\,\ref{fjarp2}. This leads to the lowest possible total photoelectron excitation vectors $\mathbf{k}'_{||}$, $-\mathbf{k}'_{||}$,
\begin{eqnarray}\label{ejap5}
\mathbf{k}_{||}'&=&-\mathbf{k}_{||}+\mathbf{k}_{CBF_{||}}+\mathbf{k}_{S_{||}},\\
&&\hspace{-1.5cm}(\mathbf{k}'_{||}, -\mathbf{k}'_{||}\,\,\,\mathrm{paired}\,\,\mathrm{photoelectrons})\nonumber\\
\label{ejap5'}-\mathbf{k}_{||}'&=&\mathbf{k}_{||}-\mathbf{k}_{CBF_{||}}-\mathbf{k}_{S_{||}}
\end{eqnarray}
taking into account that the total pair impulse is zero. The transformations of the ground state to the
excited band states from which the photoelectron emissions occur are depicted in Fig.\,\ref{fjarp2}.
However, the non-renormalized excited states are two-dimensional, thus, the states have to be extended into a
two-dimensional representation. The probabilistic electron mobility in $\bot$ direction is, however, zero which
leads to an infinite effective mass, $m_{\bot}$=$\infty$. Thus, the two-dimensional extension of the ground and
excited states reflects merely a new state of information. Since every fourth one-dimensional b-hole state in
$x$ direction in Fig.\,\ref{fjarp1} is identical, a reciprocal lattice vector
$\frac{1}{4}\mathbf{G}_{o_{\bot}}$=$\frac{1}{4}\frac{2\pi}{a}$ results. In this case, the coordinate systems of
the ground and excited states are identical  which leads to
$\mathbf{G}^{cr}_{gr_{\bot}}$=$\mathbf{G}^{cr}_{ex_{\bot}}$=$\frac{1}{4}\mathbf{G}_{o_{\bot}}$=$\frac{1}{4}\frac{2\pi}{a}$.
Thus, no additional band transformations occur for the $\mathbf{k}_{\bot}$ vectors, as opposed to the parallel
wave vectors. Altogether, the band states which occur in photoelectron emission are defined by
Eqs.\,(\ref{ejap6}) and (\ref{ejap7}).
\begin{eqnarray}\label{ejap6}
&&\hspace{-1.5cm}\mathbf{k}=(k_{\bot},k'_{||})\,\,\, \mathrm{and} \,\,\,\mathbf{k}'=(k_{\bot},-k'_{||})\\
\label{ejap7}&&\hspace{-1.5cm}k_{\bot}\,\epsilon\,[-\frac{\pi}{4a},\frac{\pi}{4a}]\hspace{1.75cm}\mathrm{first\, Brillouin\, zone}\\
\label{ejap8}\mathbf{R}_{n,m}&=&(n\cdot4a,m\cdot b)\hspace{0.5cm}\mathrm{involved\, lattice\, sites}\\
\label{ejap9}m_{||}&\simeq&\infty\hspace{2.4cm}(\mathrm{for \,the \,ground\, state})\\
\label{ejap10}m_{\bot}&=&\infty
\end{eqnarray}
In $||$ direction the overlap integrals are very small based on the doubling of the lattice constants
($2a$\,or\,$2b$). This fact and the locally pinned collective b-hole state lead to a very large effective mass
in the ground state (Eq.\,(\ref{ejap9})).\\
 $\indent$The b-hole states in $I_{y}$ and $I'_{y}$ generate negative
energy shifts. On the one hand, this is caused by the total b-hole$-$b-hole correlation energy
$E_{C_{tot}}$=$E_{C}+E'_{C}$$\lesssim$$2\cdot E_{C}\simeq-160$ meV per b-hole and additionally by the hole
self-energy $E_{S}$$\simeq-350$ meV.\cite{rjarp1} This leads to an energy shift of
$\Delta\varepsilon$=$\frac{1}{3}(E_{C_{tot}}$+$E_{S})$ for every delocalized electronic state in $I_{y}$ and $I'_{y}$.
Consequently, the Fermi energy $\varepsilon_{F}$ of the b-hole states is lowered by
$\varepsilon_{F}$=$\frac{1}{3}(E_{C_{tot}}$+$E_{S})$ relative to the hole free states, if variations of the
coulomb interactions are neglected. The b-hole bands $I_{y}, I'_{y}$ are well conducting based on the
possibility of low-energetic excitations but the hole free states $\bar{I}_{y},\bar{I}'_{y}$ are isolating.\cite{rjarp3} Thus, $\varepsilon_{F}$=$\frac{1}{3}(E_{C_{tot}}$$+E_{S})$ lines up upon the contact potential in
the ARPES experiments. If the two "photo"-holes that are left behind after the electron pair excitation were
infinitely extended in real space along the ground states of the bands $I_{y}$ or $I'_{y}$, the electrons would appear to be emitted from this
extended electronic hole state, which is defined by $\varepsilon_{F}$. However, the electronic excitations remain
locally confined based on Eqs.\,(\ref{ejap9}) and (\ref{ejap10}) as long as a coupling of the photoelectrons to the ground state
exists. Hence, the electron pair emissions initially occur with vertical kinematics depicted by the red
circles in Fig.\,\ref{fjarp1}. If the coupling to the ground state is lost and the two electrons attain the
vacuum state the paired electrons are emitted corresponding to their $\mathbf{k}$ states of Eq.\,(\ref{ejap6}).
Based on the initial local photoelectron excitations, the two photo-holes  will not be infinitely extended but
remain coupled to the photoelectrons. This leads to finite photo-hole densities at particular lattice sites
$\emph{i}$ and, therefore, to finite hole self energies $\Delta E_{S_{i}}$$<$0  (shake-up effect). Additionally,
a small perturbation of the b-hole$-$b-hole correlation energy has to be expected leading to a positive energy
shift in relation to $|E_{C_{tot}}|$ ({shake-off effect}). Overall, a negative energy shift $-\Delta_{b}$
relatively to the Fermi level $\varepsilon_{F}$ will occur
\begin{eqnarray}\label{ejap11}
\hspace{-0.5cm}{\varepsilon}_{\mathbf{k}}&=&\varepsilon_{F}+K_{S}\cdot E_{S}+K_{C_{tot}}\cdot |E_{C_{tot}}|=\varepsilon_{F}-\Delta_{\mathrm{b}}\hspace{0.0cm}
\end{eqnarray}
 with $K_{S},K_{C_{tot}}$$\ll$1 for $n_{h}$=1/8. Eqs.\,(\ref{ejap5})-(\ref{ejap7}),(\ref{ejap9}),(\ref{ejap10})
 and (\ref{ejap11}) are distinctly reflected in the experiments of Fig.\,\ref{fjarp3}.\\
 $\indent$So far, the b-hole structure has been assumed to be unchanged. However, there is the
 possibility of an additional electron$-$b-hole pair exchange ($h_{R}$,$h_{L}$) during photoelectron
 emission, depicted by the black and red arrows in Fig.\,\ref{fjarp1}. A complete transformation of the
 two b-holes $h_{R}$ and $h_{L}$ requires an additional excitation energy of $|E_{C_{tot}}|$ per photoelectron, and
 for an intermediary state  one obtains an additional excitation energy $\Delta\varepsilon_{\bar{\mathbf{k}}}$ of
 \begin{eqnarray}
\label{ejap13}
\Delta\varepsilon_{\bar{\mathbf{k}}}&=&|E_{C_{tot}}|\cdot\frac{4}{\pi}(\frac{3\pi}{4}-|\bar{\mathbf{k}}_{||}|)\\
\label{ejap14}&&\hspace{-2cm}\mathrm{for}\hspace{0,4cm}\mathbf{\bar{k}}_{||}\epsilon[\frac{\pi}{2},\frac{3\pi}{4}] \hspace{0,5cm}\mathrm{and\,(paired)}\hspace{0.2cm}\,-\mathbf{\bar{k}}_{||}\epsilon[-\frac{\pi}{2},-\frac{3\pi}{4}].
\end{eqnarray}
This additional energy shift may be termed as \emph{pair shake-up effect}. A complete pair shake-up excitation of $|E_{C_{tot}}|$ is accompanied by an annihilation of the b-hole relative state vector $\pm\mathbf{k}_{S_{||}}$, i.e  $\pm\mathbf{k}_{S_{||}}$=0 in Fig.\,\ref{fjarp2}, caused by a completely loss in periodicity of $h_{R}$ and $h_{L}$ relative to the existing b-hole structure. Averaging over the two limiting states which belong to different Hilbert spaces leads to the effective vectors $\pm\mathbf{k}'_{S_{||}}$=$\pm[c_{1}(\mathbf{k}_{S_{||}}$=$0) +c_{2}(\mathbf{k}_{S_{||}}$=${\pi}/{4})]$  with $c_{1}$+$c_{2}=1$. In addition, $\pm\mathbf{k}_{||}$ states at the zone boundaries will be occupied because the electron$-$b-hole pair exchanges imply coordinate transformations of $\pm\Delta\mathbf{r}_{||}$, which leave, however, the b-hole$-$b-hole center coordinate and the quantum clock-time $t'$ invariant. This implies additional phase shifts of $(\mathbf{k}_{CBF_{||}}\cdot(-\Delta\mathbf{r}_{||}))$ for the $\mathbf{k}_{||}$ state in Fig.\,\ref{fjarp1} (Eq.\,(\ref{ejap4})) and of $(-\mathbf{k}_{CBF_{||}}\cdot(\Delta\mathbf{r}_{||}))$ for the $-\mathbf{k}_{||}$ state. In Eq.\,(\ref{ejap4}) this leads to the transformed terms
\begin{eqnarray}
\label{ejap15}e^{i\mathbf{k}_{||}\cdot(\stackrel{[-]}{\mathbf{r}}\hspace{-0.0cm}'\hspace{-0.2cm}_{1}-\Delta\mathbf{r}_{||})} \cdot e^{i\mathbf{k}_{CBF_{||}}\cdot(-\Delta\mathbf{r}_{||})}\cdot\stackrel{[-]}{u}_{L,\mathbf{k}_{||_gr}}\hspace{-0.1cm}(\stackrel{[-]}{\mathbf{r}}\hspace{-0.05cm}'\hspace{-0.25cm}_{1}-\Delta\mathbf{r}_{||}\,)\nonumber\\
&&
\end{eqnarray}
and correspondingly for the $-\mathbf{k}_{||}$ state
\begin{eqnarray}
\label{ejap16}&&\hspace{-0.3cm}e^{-i\mathbf{k}_{||}\cdot(\stackrel{[-]}{\mathbf{r}}\hspace{-0.0cm}'\hspace{-0.2cm}_{2}+\Delta\mathbf{r}_{||})} \cdot e^{-i\mathbf{k}_{CBF_{||}}\cdot(+\Delta\mathbf{r}_{||})}\cdot\stackrel{[-]}{u}_{R,\mathbf{k}_{||_gr}}\hspace{-0.1cm}(\stackrel{[-]}{\mathbf{r}}\hspace{-0.05cm}'\hspace{-0.25cm}_{2}+\Delta\mathbf{r}_{||}\,).\nonumber\\
&&
\end{eqnarray}
If the states $\mathbf{k}_{||}$ and $-\mathbf{k}_{||}$ are assigned to the zone boundaries, ${\mathbf{k}_{||}=-\frac{\mathbf{k}_{CBF_{||}}}{2}}$ and  $-\mathbf{k}_{||}=\frac{\mathbf{k}_{CBF_{||}}}{2}$, Eqs.\,(\ref{ejap15}) and (\ref{ejap16}) may be transformed acording to
\begin{eqnarray}
\label{ejap17}(\ref{ejap15})&\equiv& e^{-i\frac{\mathbf{k}_{CBF_{||}}}{2}\cdot(\stackrel{[-]}{\mathbf{r}}\hspace{-0.0cm}'\hspace{-0.2cm}_{1}+\Delta\mathbf{r}_{||})}\cdot\stackrel{[-]}{u}_{L,\mathbf{k}_{||_gr}}(\stackrel{[-]}{\mathbf{r}}\hspace{-0.05cm}'\hspace{-0.25cm}_{1}-\Delta\mathbf{r}_{||}\,),\hspace{0.8cm}\\
\label{ejap18}(\ref{ejap16})&\equiv& e^{i\frac{\mathbf{k}_{CBF_{||}}}{2}\cdot(\stackrel{[-]}{\mathbf{r}}\hspace{-0.0cm}'\hspace{-0.2cm}_{2}-\Delta\mathbf{r}_{||})}\cdot\stackrel{[-]}{u}_{R,\mathbf{k}_{||_gr}}(\stackrel{[-]}{\mathbf{r}}\hspace{-0.05cm}'\hspace{-0.25cm}_{2}+\Delta\mathbf{r}_{||}\,)
\end{eqnarray}
In comparison to the initial states, Eqs.\,(\ref{ejap17}) and (\ref{ejap18}) represent an exchange of the relative coordinates $\pm\frac{\Delta\mathbf{r}_{||}}{2}$ of the $\mathbf{k}_{||}$ and $-\mathbf{k}_{||}$ states, i.e. relative coordinates related to a $\stackrel{[-]}{\mathrm{Cu}}$ site,  which define a symmetric pair state  $\stackrel{[-]}{\psi'}_{\mathbf{k}_{||_{gr}}}$$\cdot\,$$\stackrel{[-]}{\psi'}_{-\mathbf{k}_{||_{gr}}}$ with respect to this relative coordinate exchange.\cite{rjarp13} The transformations from the ground states to the excited states according to the transition from Eq.\,(\ref{ejap4}) to Eq.\,(\ref{ejap4'}) remain identical because the assignment of the coordinates  $(\stackrel{[-]}{\mathbf{r}}\hspace{-0.1cm}'\hspace{-0.3cm}_{1}+\Delta\mathbf{r}_{||})$ and $(\stackrel{[-]}{\mathbf{r}}\hspace{-0.1cm}'\hspace{-0.3cm}_{2}-\Delta\mathbf{r}_{||})$ to the corresponding probabilistic coordinates in the excited states $\mathbf{{r}}'_{1}$ and $\mathbf{{r}}'_{2}$ with $\mathbf{G}^{cr}_{{ex}_{||}}$=$\mathbf{G}^{cr}_{{o}_{||}}$ are indistinguishable from the initial states $\stackrel{[-]}{\mathbf{r}}\hspace{-0.1cm}'\hspace{-0.3cm}_{1}$ and $\stackrel{[-]}{\mathbf{r}}\hspace{-0.1cm}'\hspace{-0.3cm}_{2}$ , i.e. the additional coordinate shifts $\pm\Delta\mathbf{r}_{||}$ are not discernible. Thus, the additional electron$-$b-hole pair exchanges do not query the stationarity of the emission process, if the photoemissions occur from states at the zone boundaries. The population of states at the zone boundaries raises no problems. The non-dispersive nature of the $\mathbf{k}_{||}$ states within a strongly phase correlated many electron state allows a free exchange of particular $-\mathbf{k}_{||}, \mathbf{k}_{||}$ paired wave vectors.
Both effects, i.e. the formation of state vectors $\pm\mathbf{k}'_{S_{||}}$=$\pm[c_{1}(\mathbf{k}_{S_{||}}$=$0) +c_{2}(\mathbf{k}_{S_{||}}$=${\pi}/{4})]$ and the population of states at the zone boundaries, lead to wave vectors $\pm\bar{\mathbf{k}}_{||}$ according to Eq.\,(\ref{ejap14}), which are depicted in Fig.\,\ref{fjarp2}. These conclusions and the disappearance of $\pm\mathbf{k}_{||}$ states within $[0,\pm\frac{\pi}{2}]$ are clearly reflected by all experiments\cite{rjarp5}$^{-}$\cite{rjarp8} as shown, for example, in Fig.\,\ref{fjarp3}.\\
$\indent$Finally, a few words should be spent on the fact that the photoelectron pair emissions here proposed will effectively result from a two photon excitation. The existence of a phase rigid many electron pair condensate given by the two paired electron channels within every one-dimensional b-hole band $I_{y}$, $I'_{y}$ provides an explanation. This state enables the absorbtion of a photon $E=h\nu$ (several tens of eV) by the electron pair condensate as a whole. Collective pair excitations may lead to the formation of a localized many-electron bound state in real space that can be considered to be energetically retarded within the sequence of photon excitations. Then, a second absorbed photon may stimulate a photoelectron pair emission with a synchronous absorption of the photon energy $E=h\nu$  by each of the two electron channels under conservation of the pair symmetry of the states. Again, this is based on the definite pair symmetry in $\mathbf{k}_{||}$ space and real space as well as by the strong time-quantization of the CBF states.  Beyond this rather generalized description the introduced two-photon mechanism reveals a sophisticated quantum field theoretical problem which will be tackled in a separate article.
\section{Conclusion\label{secnj5}}
$\indent$The unusual electron dispersion found in single layer cuprates can be consistently explained by the TR theory of the quantum states of the $\mathrm{CuO_{2}}$ planes with a nominal hole concentration of $n_{h}=1/8$ holes/copper. Thus, these results support the conclusion of condensed electronic pair states generally existing under hole doping, already above $T_{c}$. The TR theory exhibits several new features of quantum states which are causally related to the renormalizations of the electronic states. Disjoint basis representations of the ground and excited states is one of these new quantum features including transformations of the $\mathbf{k}$ space in electronic excitations which are experimentally distinctly reflected. In addition, the existence of paired topological hole orders allows a pair shake-up effect in photoelectron emissions which explains the abrupt transition from the non-dispersive to the dispersive electronic behaviour. It is concluded that photoelectron pair emissions generally exist, at least for states near the Fermi level, which make possible a pair shake-up effect in the first place. Thus, experiments which permit a time-coincident detection of the two emitted photoelectrons in ARPES experiments would allow a direct proof of the  paired photoelectron emissions as proposed in this paper.


\begin{thebibliography}{99}
\bibitem{rjarp20}{A. Damascelli, Z. Hussain, and Z.-X. Shen, Rev. Mod. Phys. 75, 473 (2003).}
\bibitem{rjarp5}{X.J. Zhou, P. Bogdanov, S.A. Kellar, T. Noda, H. Eisaki, S. Uchida, Z. Hussain and Z.-X. Shen, Science 286, 268 (1999).}
\bibitem{rjarp6}{A. Ino, C. Kim, M. Nakamura, T. Yoshida, T. Mizokawa, Z.-X. Shen, A. Fujimori,  T. Kakeshita, H. Eisaki and S. Uchida, Phys. Rev. B 62, 4137 (2000). }
\bibitem{rjarp7}{A. Ino, C. Kim, M. Nakamura, T. Yoshida, T. Mizokawa, A. Fujimori, Z.-X. Shen, T. Kakeshita, H. Eisaki and S. Uchida, Phys. Rev. B 65, 094504 (2002). }
\bibitem{rjarp8}{T. Sato \emph{et al.}, Phys. Rev. B 64, 054502 (2001).}
\bibitem{rjarp1}{W. Winkler and K. Winkler, Physica C 450, 1 (2006).}
\bibitem{rjarp2}{W. Winkler and K. Winkler, Physica C 457, 1 (2007).}
\bibitem{rjarp9'}{S.A. Kivelson, I. P. Bindloss, E. Fradkin, V. Oganesyan, J.M. Tranquada, A. Kapitulnik and C. Howald, Rev. Mod. Phys. 75, 1201 (2003).}
\bibitem{rjarp9}{M. Ogata and H. Fugkuyama, Rep. Prog. Phys. 71, 036501 (2008).}
\bibitem{rjarp10}{P.A. Lee, Rep. Prog. Phys. 71, 012501 (2008).}
\bibitem{rjarp11}{M. Vojta, Adv. Phys. 58, 699 (2009).}
\bibitem{rjarp12}{V. Barzykin and D. Pines, Adv. Phys. 58, 1 (2009).}
\bibitem{rjarp3}{W. Winkler, J. Phys.: Conf. Ser. 153, 012034 (2009).}
\bibitem{rjarp13}{If periodic topological hole states exist, the quantum states of the valence bands form singlet pair states where the antisymmetry of the pair wave function is realized by exchanges of the spin states. In the TR theory, the antiferromagnetic state is generated by energetic deeper lying bands.}
\end{thebibliography}
\end{document}